\documentclass[conference]{IEEEtran}

\usepackage{graphicx}             
\usepackage[caption=false]{subfig}
\usepackage{url}                  
\usepackage[T1]{fontenc}
\usepackage{nicefrac}             
\usepackage{indentfirst}          
\usepackage[super, negative]{nth}
\usepackage{amsfonts}
\usepackage{multirow}
\usepackage{balance}
\usepackage{bytefield}
\usepackage{paralist}
\usepackage{cellspace}
\usepackage{algorithm}
\usepackage[noend]{algpseudocode}
\usepackage{listings}
\usepackage{subfloat}
\usepackage{tikz}
\usetikzlibrary{positioning,decorations.pathreplacing,calc,fit, shapes.geometric}
\usepackage{mathtools}
\usepackage{enumitem}
\usepackage{calc}
\usepackage{arydshln}
\usepackage[export]{adjustbox}

\usepackage{xspace}
\usepackage{array,booktabs}
\usepackage{latexsym}
\usepackage{pifont}
\usepackage{colortbl}
\usepackage{lipsum}
\usepackage{color}
\usepackage{xcolor}
\usepackage{pifont}

\definecolor{MACRouge}{RGB}{255,38,0}
\definecolor{MACBleu}{RGB}{4,51,255}
\definecolor{MACVert}{RGB}{0,143,0}
\definecolor{MACOrange}{RGB}{255,147,0}
\definecolor{MACMagenta}{RGB}{255,64,255}
\definecolor{MACAsperge}{RGB}{146,144,0}
\definecolor{MACGris}{RGB}{169,169,169}
\definecolor{MACBordeau}{RGB}{148,23,81}

\lstdefinestyle{yaml}{
    basicstyle=\color{MACBleu}\footnotesize,
    rulecolor=\color{black},
    frame = single,
    string=[s]{'}{'},
    stringstyle=\color{MACBleu},
    comment=[l]{:},
    commentstyle=\color{black},
    morecomment=[l]{-},
    linewidth=0.95\columnwidth
 }

\newcommand{\dfn}[1]{\textit{#1}}            

\newcommand{\ipsix}{\textsc{Ipv6}\xspace}
\newcommand{\ipfour}{\textsc{Ipv4}\xspace}
\newcommand{\traceroute}{\texttt{traceroute}\xspace}

\newcommand{\ping}{\texttt{ping}\xspace}
\newcommand{\mstg}{\textsc{Mstg}\xspace}
\newcommand{\mstgLong}{\textsc{\dfn{\textbf{M}icro\textbf{S}ervices \textbf{T}opology \textbf{G}enerator} (\mstg)}\xspace}

\newcommand{\tcp}{\texttt{TCP}\xspace}
\newcommand{\syn}{\texttt{SYN}\xspace}
\newcommand{\synack}{\texttt{SYN+ACK}\xspace}

\newcommand{\ack}{\texttt{ACK}\xspace}
\newcommand{\ioam}{\textsc{Ioam}\xspace}
\newcommand{\oam}{\textsc{Oam}\xspace}
\newcommand{\microservice}{microservice\xspace}
\newcommand{\microservices}{microservices\xspace}
\newcommand{\otelem}{\texttt{OpenTelemetry}\xspace}
\newcommand{\http}{\texttt{HTTP}\xspace}
\newcommand{\https}{\texttt{HTTPS}\xspace}
\newcommand{\tls}{\texttt{TLS}\xspace}
\newcommand{\jaeger}{\texttt{Jaeger}\xspace}

\newcommand{\apm}{\texttt{APM}\xspace}
\newcommand{\apmLong}{Application Performance Monitoring (\apm)\xspace}
\newcommand{\iproute}{\texttt{iproute2}\xspace}
\newcommand{\tc}{\texttt{tc-netem}\xspace}
\newcommand{\yaml}{\textsc{Yaml}\xspace}
\newcommand{\kube}{Kubernetes\xspace}
\newcommand{\dc}{Docker Compose\xspace}

\newcommand{\mtu}{Maximum Transmission Unit (\textsc{mtu})}

\begin{document}

\title{\mstg: A Flexible and Scalable Microservices Infrastructure Generator}

\author{\IEEEauthorblockN{Emilien Wansart\IEEEauthorrefmark{1},
		Maxime Goffart\IEEEauthorrefmark{2}, Justin Iurman\IEEEauthorrefmark{3} and
		Benoit Donnet\IEEEauthorrefmark{4}}
	\IEEEauthorblockA{Montefiore Institute, Universit\'e de Li\`ege, Belgium\\
		\IEEEauthorrefmark{1}emilien.wansart@uliege.be,
		\IEEEauthorrefmark{2}maxime.goffart@uliege.be,
		\IEEEauthorrefmark{3}justin.iurman@uliege.be,
		\IEEEauthorrefmark{4}benoit.donnet@uliege.be}}

\maketitle

\begin{abstract}
The last few years in the software engineering field has seen a paradigm shift from monolithic application towards architectures in which the application is split in various smaller entities (i.e., \microservices) fueled by the improved availability and ease of use of containers technologies such as Docker and Kubernetes. Those \microservices communicate between each other using networking technologies in place of function calls in traditional monolithic software. In order to be able to evaluate the potential, the modularity, and the scalability of this new approach, many tools, such as \microservices benchmarking, have been developed with that objective in mind. Unfortunately, many of these tend to focus only on the application layer while not taking the underlying networking infrastructure into consideration.

In this paper, we introduce and evaluate the performance of a new modular and scalable tool, \mstgLong, that allows to simulate both the application and networking layers of a \microservices architecture. Based on a topology described in \yaml format, \mstg generates the configuration file(s) for deploying the architecture on either \dc or \kube. Furthermore, \mstg encompasses telemetry tools, such as \apmLong relying on \otelem.  This paper fully describes \mstg, evaluates its performance, and demonstrates its potential through several use cases.
\end{abstract}

\section{Introduction}\label{intro}
The last fifteen years have witnessed a strong evolution of the Internet: from a hierarchical, relatively sparsely interconnected network to a flatter and much more densely inter-connected network~\cite{flattening,flat_model,flattening_analysis} in which hyper giant distribution networks (HGDNs, - e.g., Facebook, Google, Netflix) are responsible for a large portion of the world traffic~\cite{netflix}.  HGDNs are becoming the de-facto main actors of the modern Internet.  The very same set of actors have fueled the move to very large data center networks (DCNs), along with the evolution to cloud native networking.

In parallel, throughout the years, multiple \dfn{Operations, Administration, and Maintenance} (\oam) tools have been developed, for various layers in the protocol stack~\cite{rfc7276}, going from basic \traceroute to Bidirectional Forwarding Detection (BFD~\cite{rfc5880}) or recent UdpPinger~\cite{udppinger} and Fbtracert~\cite{fbtracert}. The measurement techniques developed under the \oam framework have the potential for performing fault detection, isolation, and performance measurements.

Telemetry information (e.g., timestamps, sequence numbers, or even generic data such as queue size and geolocation of the node that forwarded the packet) is key to HGDNs, DCNs, and Internet operators in order to tackle two particular challenges.  First, the network infrastructure must be running all the time, even in the presence of (unavoidable) equipment failure, congestion, or change of traffic patterns.  Said otherwise, it means that HGDNs and DCNs must carefully engineer their network infrastructure to be able to ensure that issues are responded to within seconds.  Network monitoring and measurements are thus of the highest importance for HGDNs and DCNs, although the available tools and methods~\cite{udppinger,fbtracert} have not kept up with the pace of growth in speed and complexity.  Second, customers want to enjoy their content in whatever context they access it: at home on one or more devices behind a DSL gateway, on a mobile device in public transportation, etc.  In addition, customers want to experience their content with the highest possible quality and the lowest delay without interfering with the network.  Consequently, HGDNs, DCNs, and classical Internet operators must carefully engineer their network to ensure the highest Quality-of-Experience (QoE) on the user side, especially with the emergence of \dfn{\microservices}~\cite{microservices_archi}, i.e., autonomous services deployed independently, with a single and clearly defined purpose~\cite{microservices_def}.

Modern cloud-native applications rely on \microservices. A single request in an application can invoke a lot of \microservices interacting with each other over the network. Consequently, it is becoming increasingly difficult to monitor and isolate a problem, e.g., a slowdown of a service. This is why Application Performance Management (\apm, based on distributed tracing tools, e.g., \otelem~\cite{opentelemetry} combined with \jaeger~\cite{jaeger} for visualization) is useful. It provides a way to observe and understand a whole chain of events in a complex interaction between \microservices. However, such \apm are not useful when the problem is not application related but rather located at the network level. Therefore, there is a need for new telemtry tools that would be able to monitor both the \microservices and the network layer underneath them.  Prior to be developed, such tools must be carefully evaluated in a controlled environment to ensure they will not disturb the various \microservices.  However, existing \microservices benchmark generators~\cite{dsb,hydragen,usuite,cloudsuite} totally ignore network resources (e.g., routers) required to run \microservices in the cloud and are not dedicated to test integrated telemetry solutions for \microservices.

In this paper, we introduce the \mstgLong, an application able to generate a complete \microservice topology that includes network components, such as routers, and \microservices.  In particular, this paper makes the following contributions:
\begin{itemize}
  \item we carefully describe \mstg.  In a nutshell, \mstg simulates \microservices and network architecture through \dfn{Docker} based on a \yaml configuration file.  The resulting topology can be deployed on a single machine with \dc~\cite{docker_compose} or distributed with \kube~\cite{kubernetes};
  \item \mstg encompasses existing telemetry solutions, such as \otelem~\cite{opentelemetry}, \jaeger~\cite{jaeger}, and In-situ \oam (\ioam -- \ioam gathers telemetry and operational information along a path, within packets, as part of an existing -- possibly additional-- header in \ipsix)~\cite{rfc9197}.  Those solutions are the basis for developing advanced telemetry tools for \microservices infrastructures;
  \item we evaluate the performance of \mstg and show it is flexible and scalable;
  \item we demonstrate, through two use cases, which are existing architecture replication and intelligent \microservice selection, suitable usages of \mstg for evaluating the impact of telemetry tools in \microservices infrastructures;
\end{itemize}

The remainder of this paper is organized as follows: Sec.~\ref{mstg} discusses how we implemented \mstg and how to use it; Sec.~\ref{eval} presents \mstg performance results; Sec.~\ref{cases} demonstrates suitable \mstg usages through two use cases; Sec.~\ref{related} positions \mstg with respect to the state of the art; finally, Sec.~\ref{ccl} concludes this paper by summarizing its main achievements.

\section{MicroServices Topology Generator (\mstg)}\label{mstg}
The \mstgLong facilitates the simulation of \microservices architectures through container technologies.  It relies on a configuration file to build a topology, which is a structure of \microservices interconnected by routers that can be parameterized at both the application and network layers.

The primary objective of this tool is to offer a customizable environment for demonstration and testing purposes. Simulating scalable \microservice topologies proves valuable in various scenarios such as validating the correctness of an architecture, conducting testing and benchmarking of a topology, and assisting in the evaluation and integration of other technologies into a \microservices architecture, such as telemetry or monitoring tools, before their inclusion into a critical production environment on which a potentially non-negligible number of users are relying.

The generated topologies consist of entities spanning different layers of the OSI model.
First, the \microservices (Layer 7) are instances of the same application, responding to requests with randomized data of a specified size, and querying other \microservices before doing so.  Second, the routers (Layer 3) are on the paths between the \microservices.  Data exchange between two \microservices follows a predetermined path across the architecture based on their IP addresses.  Routers are optional, allowing the topology to be a simple mesh of interconnected \microservices, relying on a Docker bridge network~\cite{docker_bridge} for inter-connectivity when deploying for \dc.

The ability to include routers enriches the topology with an IP layer, enabling the usage of lower-layer monitoring tools such as \ioam for network telemetry~\cite{rfc9197}, or to have an architecture closer to what one would find in a DCN or a cloud network.

\subsection{Architecture}\label{mstg.archi}
\mstg design aims at providing flexibility and ease of use.  A topology is generated from a single configuration file to ensure simplicity and portability.

\begin{figure}[!t]
    \centering
    \includegraphics[width=\columnwidth]{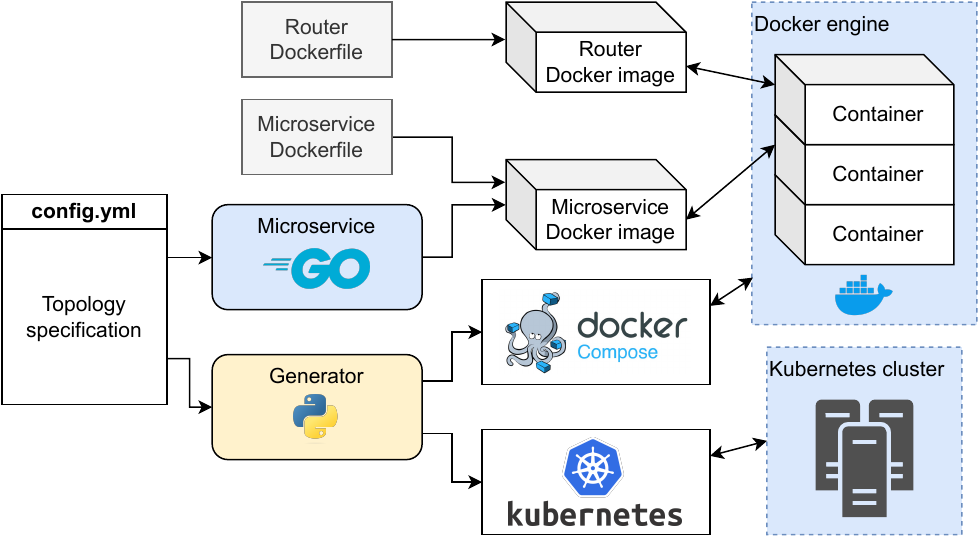}
    \caption{\mstg architecture.}
    \label{fig:architecture}
\end{figure}

\mstg architecture, depicted in Fig.~\ref{fig:architecture}, can be described as follows:
\begin{enumerate}
  \item The configuration file \texttt{config.yml} serves as the tool input, containing the topology structure with interconnected \microservices and, optionally, routers.
  \item The \textit{generator} parses the configuration file and generates either, depending on the chosen option, a configuration file for \dc when deploying on a single machine or the configuration files for a distributed deployment with \kube.
  \item The \textit{\microservice} is the application, a service responding to requests following the sequential query of other services based on the configuration file. The \microservice binary and configuration file are bundled into a Docker image.
  \item The \textit{router} Docker image is constructed exclusively from a Dockerfile. Its configuration is specified in the generated configuration file(s).
  \item Finally, the topology can be deployed on either \dc or \kube by relying on the previously generated configuration files.
\end{enumerate}

The generation process offers options, including the choice between \ipfour and \ipsix, using \http or \https for communication between the \microservices, enabling \microservices tracing (based on \dfn{\otelem}~\cite{opentelemetry} with \dfn{\jaeger}~\cite{jaeger}), and in-band telemetry with \ioam~\cite{rfc9197}.

\subsection{Components}\label{mstg.components}
\subsubsection*{Generator}
The topology generator, implemented in Python, reads the configuration file containing the description of the topology to generate the configuration file(s) for \dc or \kube by the following operations:
\begin{itemize}
	\item verification of the command line options;
	\item parsing and validation of the given configuration file;
	\item converting the configuration into the internal representation;
	\item generation of the required containers with their associated configurations to ensure the proper functioning of the topology;
	\item conversion of the internal representation into the format for the configuration files for \dc or \kube.
\end{itemize}

\subsubsection*{Microservice}
Implemented in Go, the \microservice is a piece of software that listens to HTTP or HTTPS requests and responds to them with random data of the size specified in the input file. Before responding, it queries connected \microservices sequentially with HTTP or HTTPS requests. Multiple entrypoints can be configured, each one associated with different services and having different response sizes.

The server can run over \ipfour or \ipsix and perform telemetry, if requested during the generation, by creating predefined traces and spans along with sending them to a \jaeger~\cite{jaeger} endpoint started automatically with the topology for storage and visualization.

\subsubsection*{Router}
The routers are lightweight \texttt{Alpine Linux} containers. They act as intermediaries between \microservices by forwarding their traffic according to their configured IP routes. A router is configured at the container startup through the execution of commands specified in the generated configuration file(s). These commands primarily rely on the \iproute solution tool, e.g., \texttt{ip} and \tc commands.

\subsubsection*{\jaeger}
An optional \jaeger~\cite{jaeger} container can be included in the generated topology to store and display distributed tracing data. Combined with \otelem, which is responsible for the generation of the traces in the services, \jaeger, thanks to its user interface, can be used to see the propagation of an initial request, including each subsequently generated request, across the various \microservices in the generated topology.

\subsubsection*{External Tools}
We made the choice to avoid launching one or more dedicated container(s) for simulating a client or for performing measurements on the services. Instead, the container ports are exposed to the host machine and the user is responsible for using their own tools to interact with the \microservices. Examples of such tools are \texttt{ping}, \texttt{curl}, \texttt{wrk}~\cite{wrk}, or \texttt{locust}~\cite{locust}.

\subsection{Topology Configuration}\label{mstg.config}
To construct a \microservice topology, a configuration file must be provided, which defines multiple services and routers, combined with their interactions. At least one service must be specified to have a working topology.

\begin{figure}[!t]
  \begin{lstlisting}[style=yaml]
<service_name>:
  type: service
  port: <port>
  endpoints:
    - entrypoint: <entrypoint>
      psize: <packet_size>
      connections:
        - path: <path>
          url: <url>
          [OPTIONS]

<router_name>:
  type: router
  connections:
    - path: <connection>
      [OPTIONS]
  \end{lstlisting}
  \caption{Service and router configuration template.}
  \label{lst:service}
\end{figure}

The service and router templates are illustrated in Fig.~\ref{lst:service}. The values of the fields can be described as follows:
\begin{itemize}
  \item \texttt{<service\_name>} and \texttt{<router\_name>}: the \microservice and router names;
  \item \texttt{<port>}: the port used for interacting with the \microservice;
  \item \texttt{<entrypoint>}: the URL at which the \microservice can be contacted;
  \item \texttt{<packet\_size>}: the size of the data returned by the \microservice for the specified endpoint;
  \item \texttt{connections}: describes the requests the \microservice will make to other \microservices prior to responding to incoming requests. Each connection includes a \texttt{<path>} specifying the route and a \texttt{<url>} indicating the endpoint. Additional optional parameters, denoted as \texttt{[OPTIONS]}, can be included.
\end{itemize}

Multiple endpoints and connections per endpoint can be specified. Connections towards other \microservices are specified using either:
\begin{itemize}
  \item a path when multiple hops are involved before reaching the destination. Intermediate hops, which can only be routers, are separated by the 2 characters '\texttt{->}' and the last hop must be a \microservice (e.g., \texttt{r1->r2->s1} where both \texttt{r1} and \texttt{r2} are routers while \texttt{s1} is a \microservice);
  \item the name of another service when the two services are connected directly (e.g., \texttt{s1}).
\end{itemize}
In a router, the \texttt{<connections>} field must match one of the \texttt{<connections>} field specified in a \microservice going through this router. For matching, the next hop must be specified. For instance, \texttt{r1->r2->s2} in \microservice \texttt{s1}, \texttt{r2} in router \texttt{r1}, and \texttt{s2} in router \texttt{r2}. This linkage is required for being able to generate the commands to correctly configure the data plane in the containers.

The \texttt{[OPTIONS]} field is optional and can contain a list of network options that introduce impairments to the network. Available options include:
\begin{itemize}
  \item \texttt{mtu: <value>}: adjusting the IP \mtu;
  \item \texttt{buffer\_size: <value>}: modifying the queue size in the Linux kernel;
  \item \texttt{rate: <value>mbit}: adjusting the link rate;
  \item \texttt{delay: <value>us}: introducing delay;
  \item \texttt{jitter: <value>us}: introducing jitter;
  \item \texttt{loss: <value>\%}: adjusting the packet loss rate;
  \item \texttt{corrupt: <value>\%}: adjusting the corrupted packets rate;
  \item \texttt{duplicate: <value>\%}: adjusting the duplicated packets rate;
  \item \texttt{reorder: <value>\%}: adjusting the packet reordering rate;
  \item \texttt{timers: <list>}: modifications of one or more of the aforementioned options over time.
\end{itemize}

A timer can be declared as shown in the template Fig.~\ref{timer} by choosing a \texttt{<newValue>} for the \texttt{<option>} that will be set \texttt{<start>} seconds after the launch of the container and restore to its original value after \texttt{<duration>} seconds. This allows to simulate different network environments in which the \microservices are executing.
\begin{figure}[!t]
	\begin{lstlisting}[style=yaml]
- option: <option>
  start: <start>
  duration: <duration>
  newValue: <newValue>
	\end{lstlisting}
	\caption{Template for a timer.}
	\label{timer}
\end{figure}

\begin{figure}[!t]
	\begin{lstlisting}[style=yaml]
frontend:
  type: service
  port: 80
  endpoints:
    - entrypoint: /
      psize: 1024
      connections:
        - path: r1->db
          url: /
          rate: 100mbit
          timers:
            - option: rate
              start: 10
              duration: 30
              newValue: 1gbit
    - entrypoint: /payment
      psize: 512
      connections:
        - path: payment
          url: /
r1:
  type: router
  connections:
    - path: db
db:
  type: service
  port: 10001
  endpoints:
    - entrypoint: /
      psize: 128
payment:
  type: service
  port: 10002
  endpoints:
    - entrypoint: /
      psize: 256
	\end{lstlisting}
	\caption{Example of a configuration file.}
	\label{lst:example_config}
\end{figure}

\begin{figure}[!t]
	\centering
	\includegraphics[width=\columnwidth]{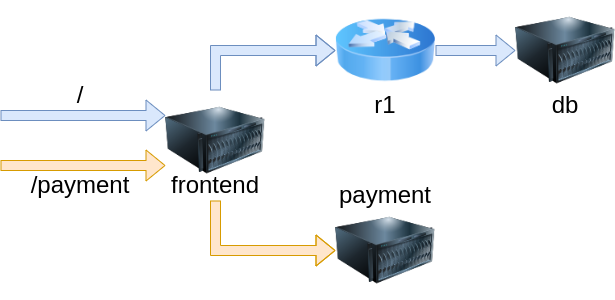}
	\caption{Architecture illustrating configuration file of Fig.~\ref{lst:example_config}.}
	\label{fig:example_config}
\end{figure}

An example of a valid configuration file is given in Fig.~\ref{lst:example_config}. This file will generate the architecture depicted in Fig.~\ref{fig:example_config}. The \texttt{frontend} service can contact the \texttt{db} service through router \texttt{r1} and the service \texttt{payment} directly. The link between \texttt{frontend} and \texttt{r1} will initially have a rate of 100Mb/s then a rate of 1Gb/s between time \textsc{t=10s} and \textsc{t=30s}. The initial rate will be restored after \textsc{t=30s}.

\subsection{Limitations}\label{mstg.limits}

\mstg first limitation is related to the sizes of \ipfour or \ipsix addresses. When generating for \dc with \ipfour, we made the arbitrary decision to use networks with a prefix of 22 bits. Thus, we have 10 bits left for the host part of the IP addresses. So, one can have up to $2^{22}-1$ (Docker) networks in the generated architecture with a maximum of $2^{10}-2$ hosts per network. On the contrary, for \ipsix, we fixed the network prefix to 64 bits, hence 64 bits for the host part. Consequently, one can simulate up to $2^{64} - 1$ (Docker) networks with up to $2^{64}$ hosts in each network.

Furthermore, there is a limitation implicated by the 16-bit size of port numbers. Following IANA~\cite{ianaports}, port numbers from 0 up to 1023 included are system ports.
Therefore, to prevent any undesired or unexpected behaviors from the system, one should refrain from using them. This leaves one with ports $\in \left[1024; 65,535\right]$ meaning that you can have up to $64,510$ services in your architecture if one wants them to be reachable from the host assuming the generation to Docker compose.

Finally, there is an upper limit on the architecture size that you can run due to computing power limit. When generating the topology for \dc, you will be limited by the hardware on a single machine. Sec.~\ref{eval} provides an evaluation of the performance in such a case. Nonetheless, this limit can be overcome with a deployment to a \kube cluster. Then, one will be restrained by the number of machines one can allocate to the cluster.

\section{Evaluation}\label{eval}
In this section, we validate and assess \mstg performance in generating \microservices topologies. We outline the methodology adopted (Sec.~\ref{eval.methodology}) and, thereafter, discuss the results (Sec.~\ref{eval.verif} and Sec.~\ref{eval.bench}).

\subsection{Methodology}\label{eval.methodology}
Our methodology is structured to comprehensively evaluate \mstg capabilities. It involves two main aspects: firstly, ensuring the correctness of the tool's functionality, where input variables are accurately reflected in the generated topologies, and secondly, evaluating the tool's performance.

We conduct \mstg scalability benchmark by examining its impact on a single host machine, including CPU and RAM usage. Additionally, we assess the scalability of the generated topology by determining the tool's ability to simulate varying numbers of \microservices and routers, considering various input variables specified in the configuration file.

Unless specified otherwise, all measurements presented here are averages obtained from 10 samples collected over a 30-second interval during 10 distinct executions of the tests to ensure representativeness of the statistics while the shaded areas around the curves are standard deviations. All the experiments were conducted using \http for communication between the \microservices to have measurements for the most basic case (i.e., \http) without the overhead introduced by \tls. The experiments were conducted on a server equipped with an Intel Xeon CPU E5-2630 v3 with a base frequency of 2.40GHz, 8 Cores, 16 Threads, 32GB of RAM, and running Linux 5.17.

\subsection{Verification}\label{eval.verif}

\begin{figure*}[!h]
	\begin{center}
		\subfloat[Maximum network utilization (RX \& TX) per packet size.]{
			\includegraphics[width=0.3\linewidth]{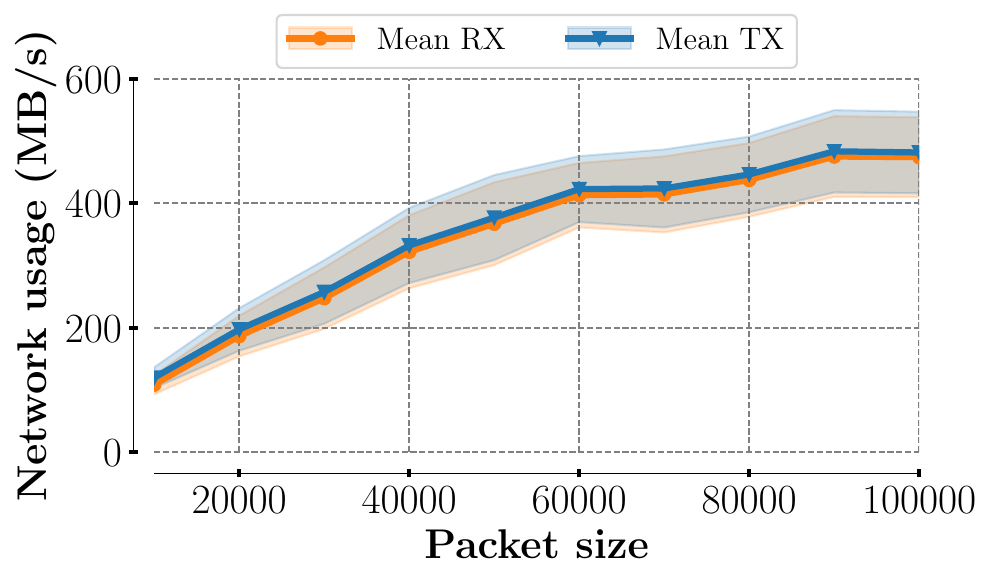}
			\label{fig:packet_size_rx_tx}
		}
		\subfloat[RTT per link delay.]{
			\includegraphics[width=0.3\linewidth]{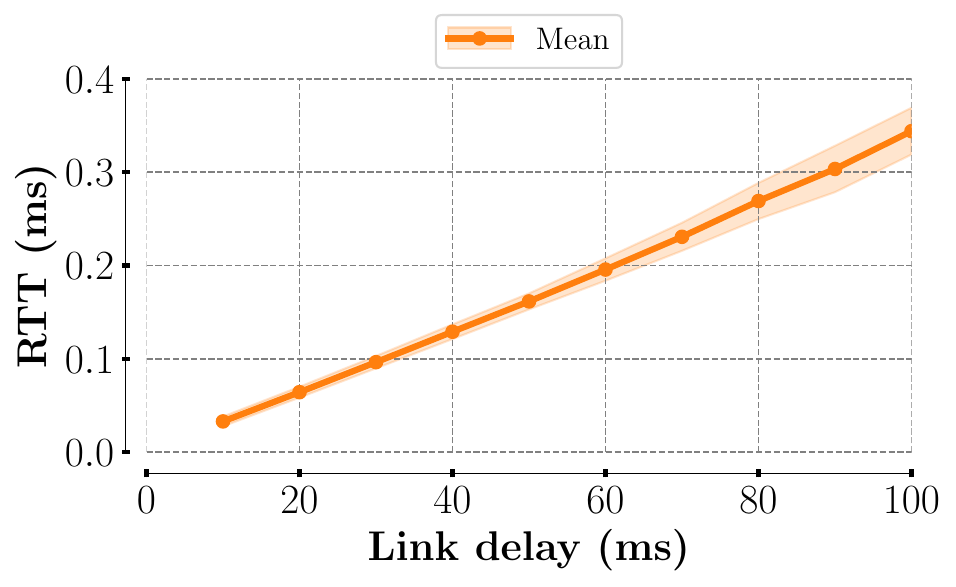}
			\label{fig:option_delay_rtt}
		}
		\subfloat[Maximum request rate per loss rate.]{
			\includegraphics[width=0.3\linewidth]{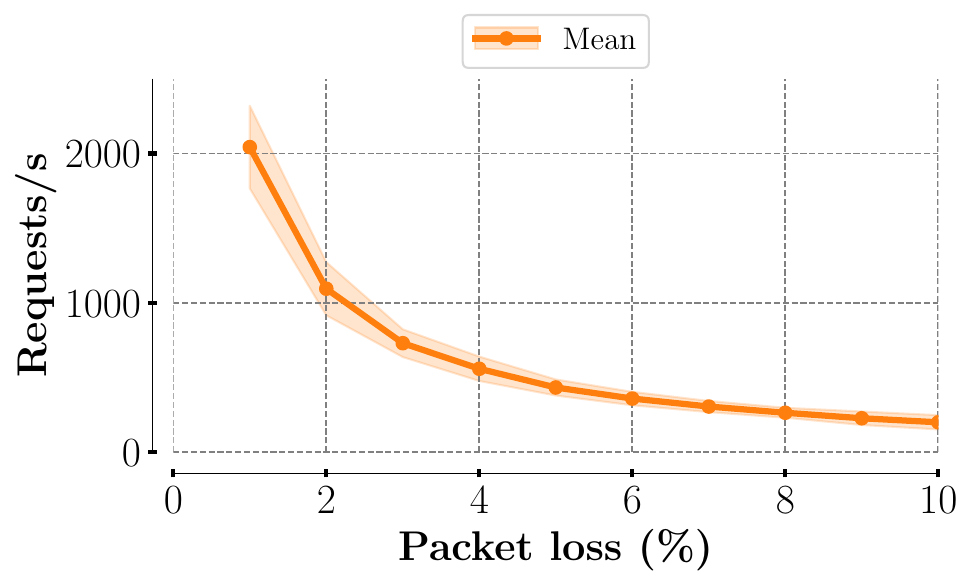}
			\label{fig:option_loss_reqs}
		}
	\end{center}
	\caption{MSTG verification.}
	\label{fig.verif}
\end{figure*}

The ability to simulate networks having real-world characteristics is dependent on Docker and the Linux Kernel since \mstg relies on them. To verify the correctness of \mstg, we focus on the following key configuration options since they have substantial performance impacts that can be assessed by performance measurements: \microservice packet size, router link delay, and loss rate. We selected the values of the variables presented hereafter in order to cover real deployments of \microservices architecture. The configuration options concerning the \microservices and routers connectivity can be verified visually using the \jaeger UI.

For the \microservices, we examine the impact of the packet size on network utilization, as shown in Fig.~\ref{fig:packet_size_rx_tx}. The graph indicates the influence of the packet size on the maximum network data input (RX) and output (TX) for a simple topology with two \microservices connected by a router. As depicted, the network usage is increasing as the size of packets is increasing. This is the expected result.

For the routers, we assess the impact of the link delay on the round-trip time (RTT) and the effect of loss rate on the maximum request rate attainable, as shown in Fig.~\ref{fig:option_delay_rtt} and~\ref{fig:option_loss_reqs}, the topology being identical to the previous experiment. Due to the linear relationship between the link delay specified in the topology configuration and the measured RTT along with the relation between the chosen packet loss rate in the configuration and the measured maximum number of requests per second, we can conclude that \mstg is rightly simulating a network.

\subsection{Benchmarking}\label{eval.bench}

\begin{figure*}[h]
	\begin{center}
		\subfloat[RAM usage.]{
			\includegraphics[width=0.45\linewidth]{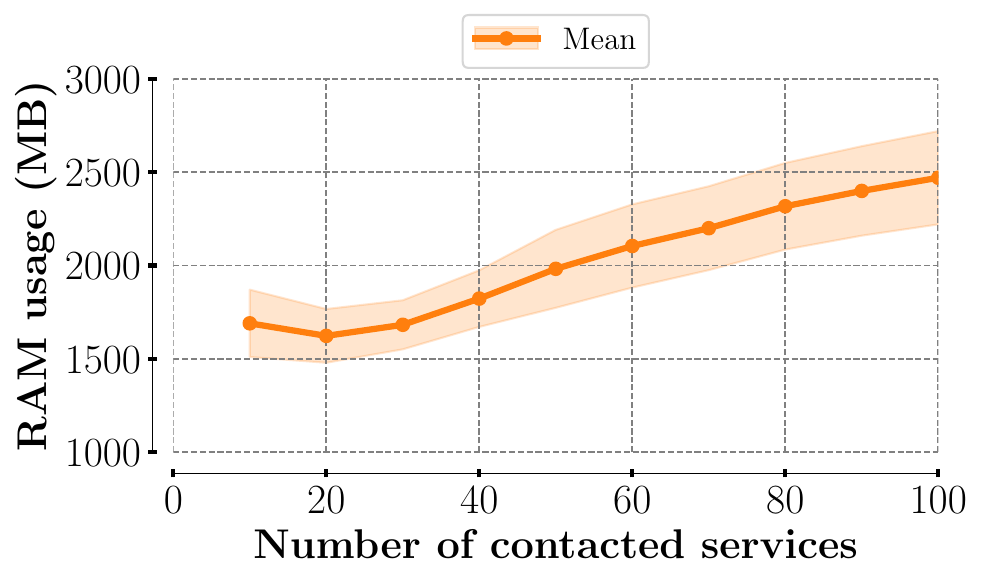}
			\label{fig:ram_usage}
		}
		\subfloat[Start and stop times per number of entities.]{
			\includegraphics[width=0.45\linewidth]{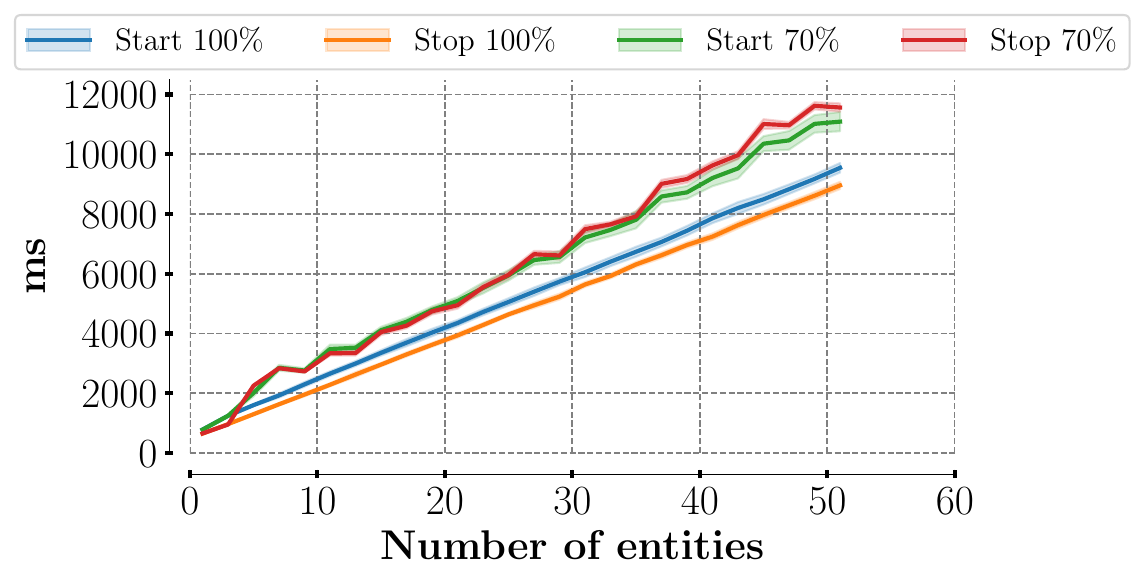}
			\label{fig:start_stop}
		}
	\end{center}
	\caption{\mstg benchmarking.}
\end{figure*}

We have evaluated the CPU and RAM usage depending on the number of entities (services or routers). We quickly reach 100\% of CPU usage while the relation between the memory usage and the number of entities is less than linear, as depicted on Fig.~\ref{fig:ram_usage}.

Furthermore, in Fig.~\ref{fig:start_stop}, to assess practicality, we present the average wall-clock time, calculated over 30 iterations, required to start and stop the architecture. We examine two scenarios: one with only services in the architecture and another with a composition of 70\% services and 30\% routers. As observed on the figure, there exists a linear relationship between the number of entities in the topology and the time needed to start and stop the architecture.

The build time is also linked to the topology size but is negligible (85ms in average over 100 iterations to generate an architecture composed of 100 services).

\begin{figure}
	\includegraphics[width=\columnwidth]{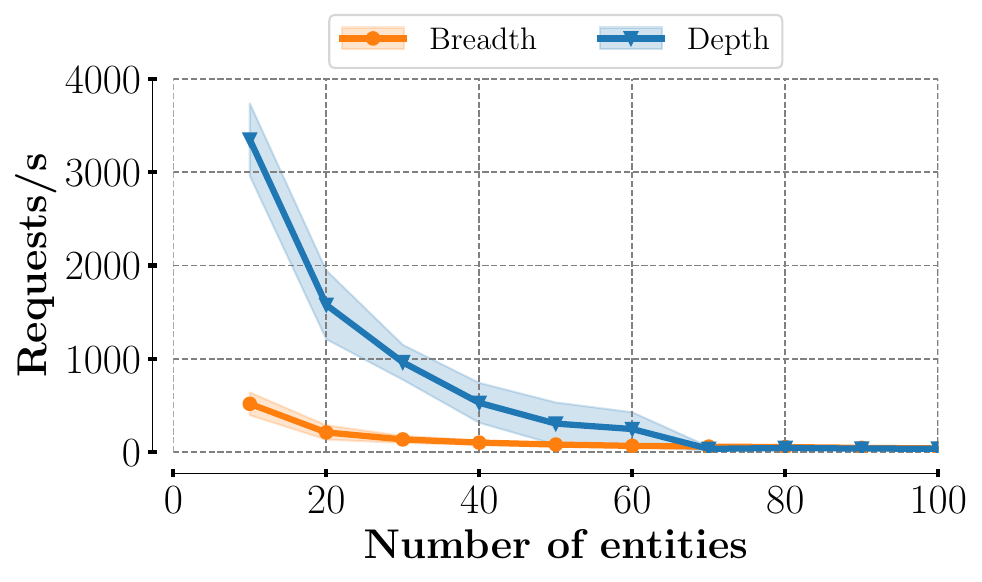}
	\centering
	\caption{Number of requests per second.}
	\label{fig:reqs}
\end{figure}

Finally, scalability is assessed by measuring the maximum number of requests possible at the frontend for topologies of increasing sizes. Two experiments differentiate the impact of \microservices and routers: one measuring the number of requests per \microservice breadth (i.e., number of \microservices queried sequentially by a frontend) and the other per router depth (i.e., number of routers between two \microservices). The results are shown in Fig~\ref{fig:reqs}. As expected, in both cases, we can see a hit in the number of requests per second. This is a normal behavior since the service needs to wait for either the responses from the other services or the propagation of the packets across all the routers. Thus, it limits the amounts of requests to which we can reply.

\section{Use Cases}\label{cases}
In this section, we explore various use cases where \mstg proves to be a valuable tool. We discuss its application in replicating existing architectures (Sec.~\ref{cases.open}) and showcase a scenario where \mstg, combined with telemetry, assists in intelligent \microservices selection (Sec.~\ref{cases.select}).

\begin{figure}[h]
	\includegraphics[width=\columnwidth]{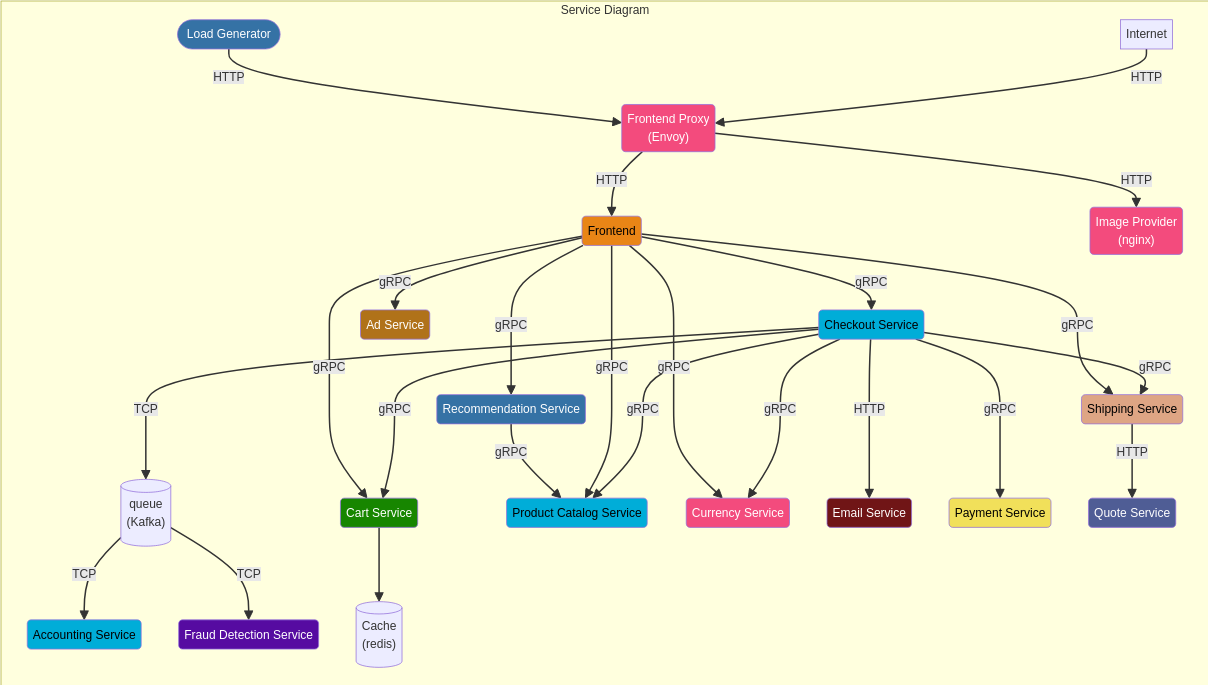}
	\centering
	\caption{Architecture of \otelem demo.}
	\label{fig:otelem_demo}
\end{figure}

\subsection{Replicating \otelem Demo Architecture}\label{cases.open}
We leverage \mstg to replicate an existing architecture from the \otelem Demo~\cite{opentelemetry_demo}, depicted on Fig.~\ref{fig:otelem_demo}. It has been designed for demonstrating \otelem's capabilities by generating telemetry data like traces, metrics, and logs.

By crafting a custom configuration file for \mstg, we recreate an architecture with the same topology as the original one. Additionally, the network options can be tuned to yield comparable performances. Thus, \mstg achieves the same goal as the original one, but with a fully configurable topology that can be expanded with customizable routers at network layer.

\subsubsection{Traces Comparison}
\begin{figure}[h]
	\begin{center}
		\subfloat[\otelem Demo.]{
			\includegraphics[width=0.45\columnwidth]{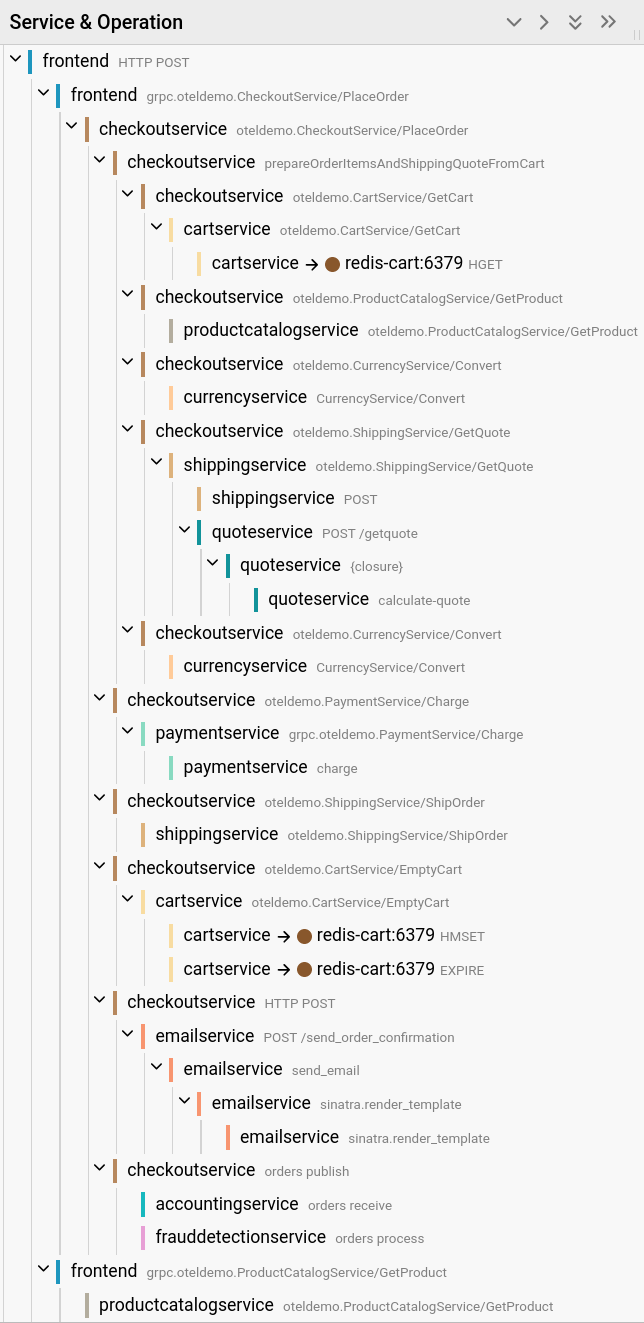}
			\label{fig:otel-demo_traces}
		}
		\subfloat[\mstg replica.]{
			\includegraphics[width=0.45\columnwidth]{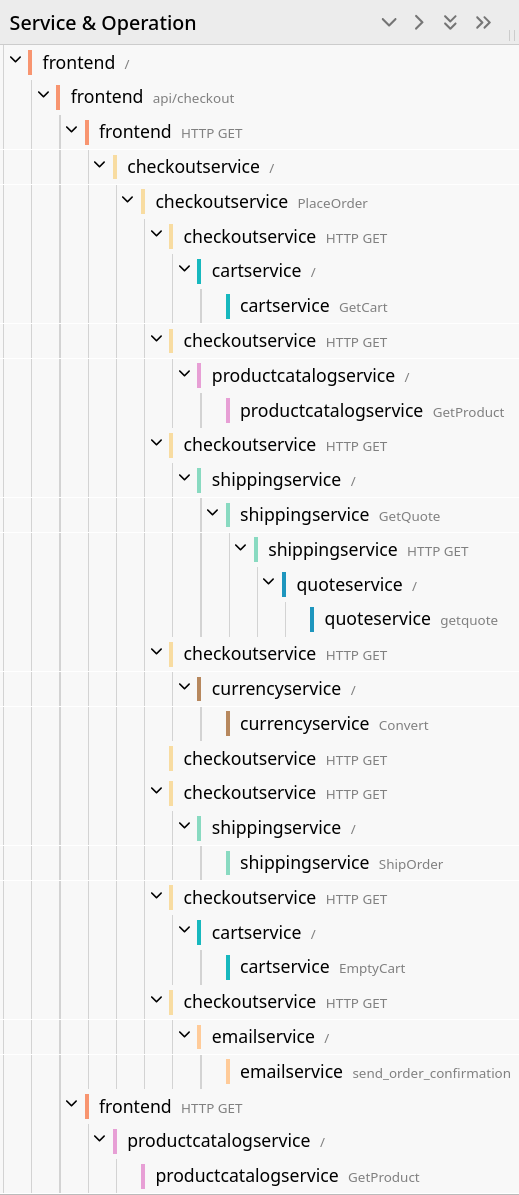}
			\label{fig:replica_traces}
		}
	\end{center}
	\caption{\otelem Demo and \mstg replica traces.}
	\label{fig:traces_comparison}
\end{figure}

We compare the traces obtained from the original otel-demo topology and its \mstg replica to validate the capability of \mstg to replicate an existing architecture. Traces are generated by querying the same endpoint on both topologies. The results, displayed in \jaeger UI as in Fig.~\ref{fig:otel-demo_traces} and Fig.~\ref{fig:replica_traces}, showcase the similarities and differences between the two.

In Fig.~\ref{fig:traces_comparison}, both topologies exhibit similar interconnections between the \microservices, with minor differences. The primary distinction lies in the arrangement of sub-spans, where the original provides more detailed information. Additionally, the replica's execution time is approximately $6$ times faster, as expected, since it only mimics the original without doing any substantial computing.

\subsubsection{Network Impairments}
The otel-demo has feature flags\footnote{Those flags allow one to enable or disable specific features in an application without changing its code.} that can trigger failures in various \microservices. \mstg replicates this mechanism, but introduces network impairments.

\begin{figure}[h]
	\includegraphics[width=\columnwidth]{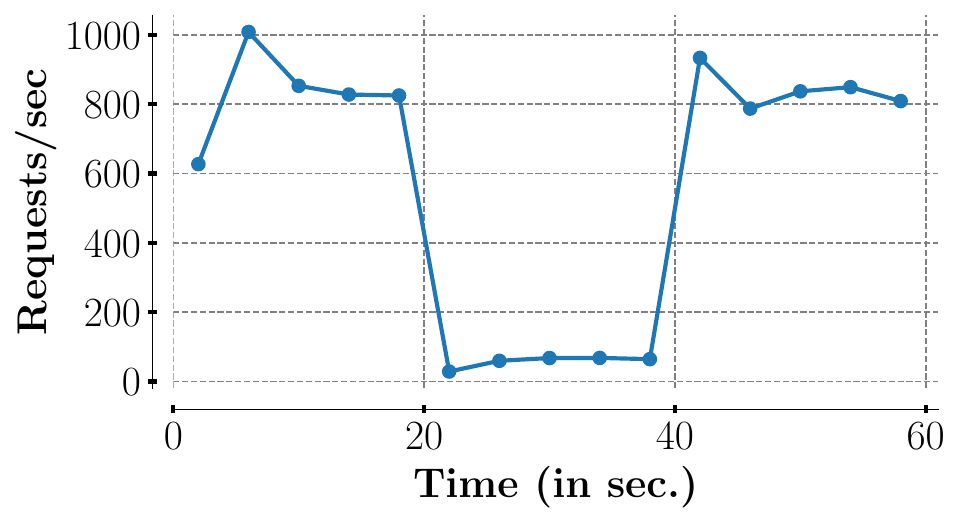}
	\centering
	\caption{Maximum request rate over time.}
	\label{fig:otel-demo_rates}
\end{figure}

Demonstrating \mstg's capability to generate failure scenarios using timers, we modify the replica. The \texttt{rate} option, governing the maximum rate of an interface, is adjusted temporarily to emulate a network link failure. The link connecting the \textit{frontend} to the \textit{checkoutservice} initially has a rate of 1 Gb/s. After 20 seconds, it is reduced to 500 Kb/s for the next 20 seconds. Fig.~\ref{fig:otel-demo_rates} illustrates the maximum \http request rate over time. As expected, a noticeable drop in the request rate occurs when the timer triggers.

\subsection{Microservice Selection}\label{cases.select}
\mstg can also be used to evaluate the possibility of providing an intelligent \microservice selection. Usually, in a \microservices architecture, there are many instances of the same service to increase the availability and provide the best experience to the end users. In such a case, the instances are behind a proxy, which is acting as a load-balancer, that redirects each request to one of the available instances based on some predefined criteria.  This section describes three scenarios.

\begin{figure}[h]
	\includegraphics[width=\columnwidth]{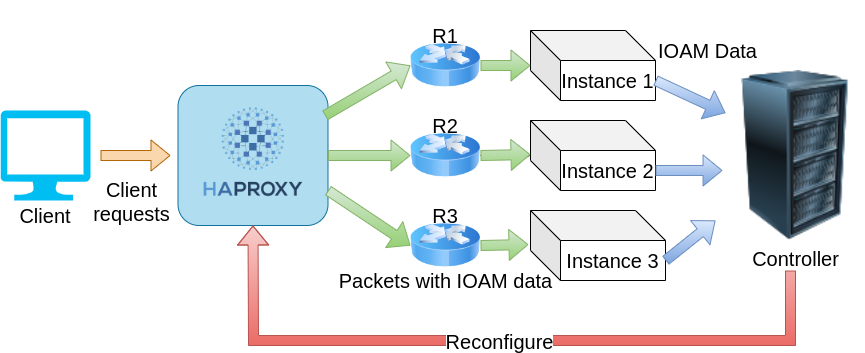}
	\centering
	\caption{Architecture of service selection with HAProxy.}
	\label{fig:service_selection_haproxy}
\end{figure}

\subsubsection{Modified \tcp Three-Way Handshake}
The best instance at a precise moment is selected by the server, acting as a proxy, by using \ioam for optimized instance selection. Then, all the subsequent traffic must go through the proxy. This is something \ping, \traceroute, or even BFD~\cite{rfc5880} cannot do, as \ioam also carries useful data and is designed for that purpose. To do so, the following steps are performed:
\begin{enumerate}
  \item The client initiates the \tcp handshake with the proxy by sending a \syn;
  \item The proxy replicates the \syn to each instance;
  \item Each instance replies to the \syn with \synack containing \ioam data, which will be collected on the return path;
  \item The proxy receives the \synack from every instance and determines the best instance based on the \ioam data extracted from each of the received \synack;
  \item The proxy sends a \synack to the client and kills the other half-open connections;
  \item The client completes the handshake with the proxy;
  \item All the subsequent traffic reach the selected instance by going through the proxy.
\end{enumerate}

\begin{figure}[h]
	\includegraphics[width=\columnwidth]{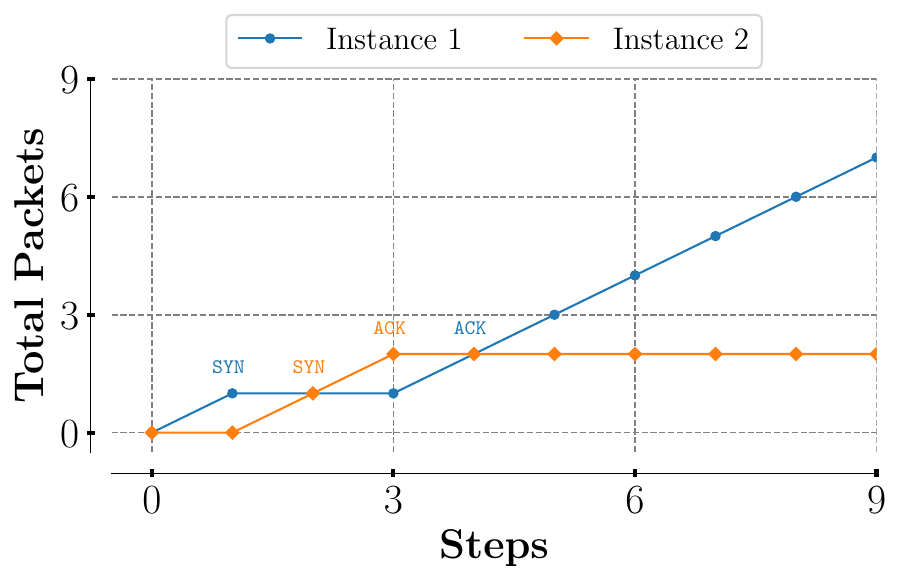}
	\centering
	\caption{Amount of packets received in each instance.}
	\label{fig:selection_tcp_graph}
\end{figure}

To ensure the proper functioning of our implementation, we tested it on an architecture composed of a client that can reach 2 instances of a service located behind a proxy where each instance is separated from the proxy by an intermediary router. We measured the total number of packets received in each instance over time as depicted in Fig.~\ref{fig:selection_tcp_graph}. As we can see, the first instance was selected by the proxy because the second instance does not receive additional packets after the \syn and \ack packets because the client interacts directly with the selected instance after the TCP handshake.

The advantages of this approach are the following. First, the client does not need to be aware of the instances and only need to communicate with the proxy. Furthermore, the proxy can benefit from the insights provided by \ioam to select the "best" instance and reconfigure itself based on the received \ioam data. Yet, in this approach, the proxy becomes a bottleneck since all the traffic must go through it. Additionally, we are gathering the \ioam data on the return path (i.e., instance $\to$ proxy) which might be different than the forward path depending on the configuration of the routers.

\subsubsection{Using Software-defined Networking (SDN) Controller}
Another solution for an intelligent service selection is to combine \ioam with HAProxy~\cite{haproxy}, which is a load-balancer. As represented in Fig.~\ref{fig:service_selection_haproxy}, a client reaches the \microservice instances through a transparent proxy. The load-balancer is responsible for sending packets augmented with \ioam data, which will contain the queue depth of every router on the path, to every instance at regular time intervals. Once the instances receive the packets, they extract the \ioam data, gathered along the hops (routers) between them and the proxy, and send it to a centralized controller. The controller will reconfigure the proxy by adjusting the distribution of the packets proportionally to the metrics observed on the paths and collected by \ioam.

\begin{figure}
	\includegraphics[width=\columnwidth]{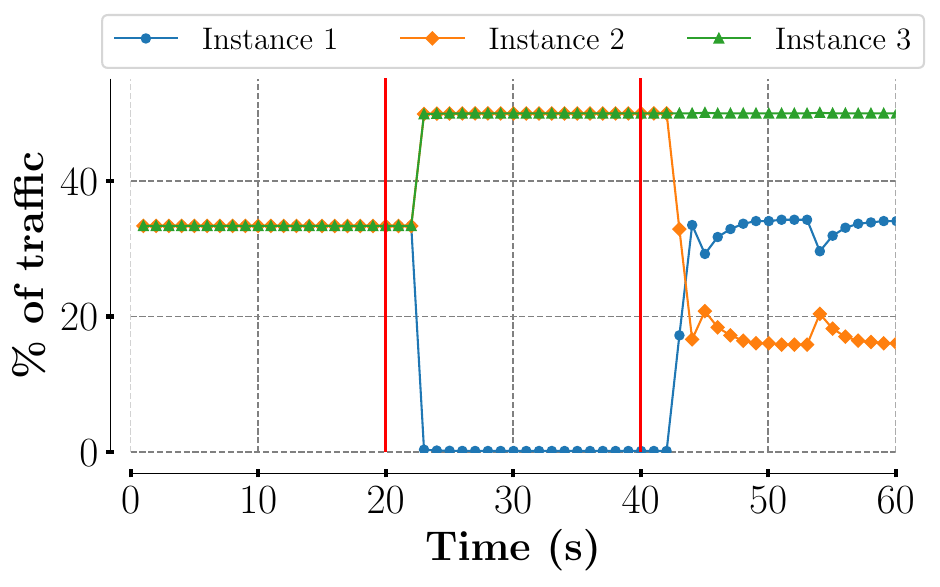}
	\centering
	\caption{Service selection with HAProxy.}
	\label{fig:haproxy_graph}
\end{figure}

We tested the version using HAProxy with the architecture depicted in Fig.~\ref{fig:service_selection_haproxy} and measured the percentage of the traffic propagated to each instance over time. At time \textsc{t=20s}, we generated a load between the load-balancer and the first instance, while at \textsc{t=40s}, we generate a load twice as high between HAProxy and the second instance. As observable in Fig.~\ref{fig:haproxy_graph}, a few seconds after the loads are generated, the collector computes the new proportions for each instance and reconfigures HAProxy with these new values. Thus, the instances on overloaded paths get less of the subsequent traffic based on the \ioam metadata.

\subsubsection{Using M-Anycast}
\begin{figure}
	\includegraphics[width=\columnwidth]{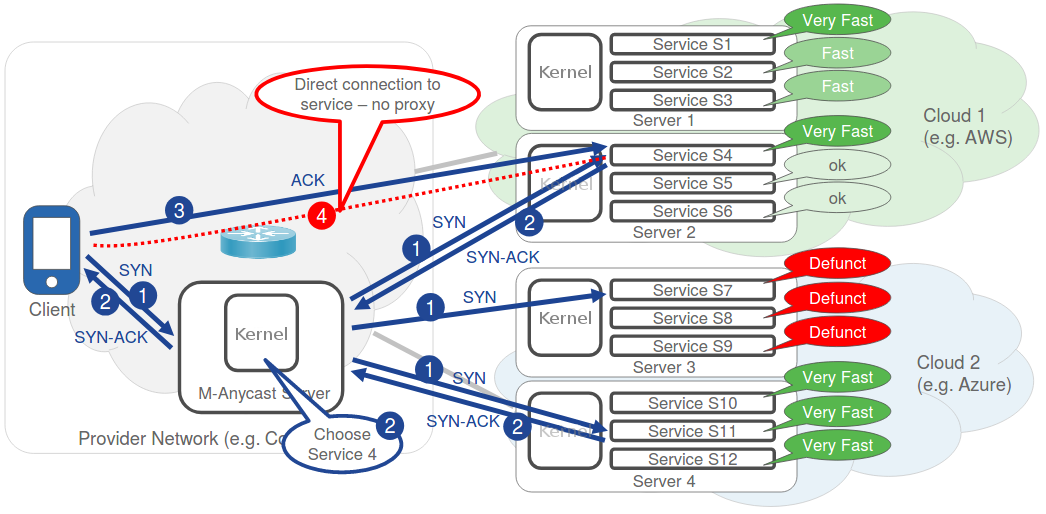}
	\centering
	\caption{Service Selection (picture from \cite{iurman_ioam}).}
	\label{fig:service_selection}
\end{figure}

An alternative would be to use \dfn{M-Anycast}~\cite{manycast}, as illustrated in Fig.~\ref{fig:service_selection}, which combines Segment Routing~\cite{rfc8986} and \ioam, so that the traffic does not go through the proxy once an instance is chosen.

The client will initiate the \tcp handshake process with the proxy, which will be responsible for selecting the ``best'' instance for the client by replicating the \syn packet to many instances and analyzing the received \synack containing \ioam data. Once the "best" instance has been chosen by the proxy, the \tcp \synack coming from it is forwarded back to the client. Then, the end-user can terminate the \tcp handshake with the selected instance and, subsequently, interact with it directly without going through the proxy.

\section{Related Work}\label{related}
Network topology generators have been developed over the last twenty years (e.g.,~\cite{brite,syntgen,realnet,igen}), following the advances in network topology discovery~\cite{survey} and modeling~\cite{vespignani,systematic,orbis}.  Those generators produce graphs trying to reproduce Internet or autonomous system properties.

Industrial network topology generators have recently emerged (e.g., GENIND~\cite{genind} or ESCALB~\cite{escalb}) and are typically designed for Internet of Things (IoT) based networks.  Those generators produce multi-level graphs, but traffic aspects are ignored, as well as potential \microservices run by IoT networks.

Closer to our work, the previous decade has seen the rise of \microservices generators for benchmarking purposes~\cite{dsb,hydragen,usuite,cloudsuite}. Death-Star Bench~\cite{dsb} exemplifies several use cases (e.g., social network or hotel reservation).  HydraGen~\cite{hydragen} generates benchmarks with different computational complexities and topologies.   $\upsilon$-suite~\cite{usuite} is used to study how OS and network overheads impact \microservice latency.  Finally, CloudSuite~\cite{cloudsuite} focuses on scale-out workloads. Other examples more explicitly focus on benchmarking particular \microservices, mainly from the e-commerce world~\cite{trainticket,bookinfo,sockshop,onlineboutique}.  All those solutions, on the contrary to \mstg, totally ignore network resources (e.g., routers), require to run \microservices in the cloud and do not encompass telemetry possibilities.

\section{Conclusion}\label{ccl}
Undoubtedly, \microservice architectures are becoming the default paradigm and can replace monolithic applications in most scenarios. Yet, most tools to evaluate different criteria about this new approach give too few considerations to the networking layer, which is a critical component for this software pattern.

This paper proposed and evaluated \mstgLong, a \microservices topology generator that gives back to the networking layer the attention it deserves by creating a modular and scalable \microservices topology generator.  \mstg offers the possibility to simulate both the networking and application layers, which are configurable by the end-users in a configuration file used as input. As demonstrated by empirical measurements, \mstg provides to users the ability to easily and rapidly test their \microservice architectures and their integration with other telemetry or monitoring technologies before deploying them on a production environment on which a significant amount of customers may depend.

\section*{Software Artefact}
Source code for \mstg and scripts used for building use cases described in this paper are available at \url{https://github.com/Advanced-Observability/Micro-Services-Topology-Generator}.

This is supported by the CyberExcellence project funded by the Walloon Region, under number 2110186, and the Feder CyberGalaxia project.

\normalsize{
	\balance

}

\end{document}